\documentclass[prd,showpacs,preprintnumbers,amsmath,amssymb,nofootinbib]{revtex4}
\everymath{\displaystyle}
\usepackage[dvips]{graphicx}
\usepackage{longtable}

\begin{document}

\preprint{TKYNT-06-5, IMSc/2005/4/11}

\title{ Is the 2-Flavor Chiral Transition of First Order?}

\author{Sanatan Digal}
\affiliation{Department of Physics, University of Tokyo,
  Tokyo 113-0033, Japan\\
Radiation Lab., RIKEN, 2-1 Hirosawa, Wako, Saitama, 351-0198,
Japan\\
and\\
Institute of Mathematical Sciences, C.I.T. Campus,
Taramani, Chennai  600 113, India}

\begin{abstract}

We study the effects of the interaction between the Chiral 
condensate and the Polyakov loop on the chiral transition within an 
effective Lagrangian. We find that the effects of the interaction
change the order of the phase transition when the explicit breaking 
of the $Z_N$ symmetry of the Polyakov loop is large. Our results
suggest that the chiral transition in 2-flavor QCD may be first 
order.

\end{abstract}


\maketitle

\section{\bf Introduction}

QCD matter with $N_f$ flavors, $N$ colors and zero baryon chemical potential
undergoes two finite temperature phase transitions, the chiral transition 
and the deconfinement. The
chiral transition is associated with the spontaneous breaking of the chiral 
symmetry, $SU(N_f)_L\times SU(N_f)_R \to SU(N_f)_V$, below the critical
temperature ($T_\chi$) for massless quarks.
The order parameter for this transition is the chiral condensate. 
On the other hand the deconfinement transition
is associated with the spontaneous symmetry breaking,
$Z_N \subset SU(N) \to \bf{1}$, 
above the critical temperature $T_d$ for infinitely heavy
quarks. The order parameter for this transition is the Polyakov loop 
expectation value. For finite non-zero quark masses both the chiral
symmetry and the $Z_N$ symmetry are explicitly broken. 
Nevertheless these two transitions
show up as crossover, second or first order transitions depending on the 
values of the quark masses. The chiral transition depends on the number 
of quark flavors and the deconfinement transition 
depends on the number of colors. Since the chiral symmetry and the $Z_N$ 
symmetry seem nothing to do with each other one would expect that these two 
transitions occur independently. However lattice QCD calculations have 
shown that both these transitions occur simultaneously, $T_\chi = T_d$
[1-5]. Furthermore strong correlation between the chiral condensate
and the Polyakov loop is observed around the phase transition
point\cite{dgl}. This is clear evidence that there is interaction
between these two order parameters. So studying the interaction between 
the chiral condensate
and the Polyakov loop is fundamental to understanding the interplay 
between the chiral transition and deconfinement. 
There are several studies on the possible causes of 
the simultaneous occurrence of the chiral transition and the deconfinement 
transition. Mixing between the gauge and matter field 
operators has been suggested to explain the simultaneous chiral and 
deconfinement transitions \cite{christian,hf}. Some other 
studies consider that for small quark masses the chiral transition
drives the deconfinement transition \cite{dgl,sannino}. 

Though most of the studies
are concerned with why $T_\chi = T_d$ only a few have considered the effect 
of the interaction on the phase transition itself \cite{midorikawa}. 
It seems natural to 
expect that if the interaction between the two order parameters can
result in the simultaneous occurrence of the two transitions then the
interaction may also have important effect on the phase transition.
One of the most interesting cases to study for the possible 
effects of this interaction is the 2-flavor chiral transition. Lattice
QCD calculations have not yet been able to settle on the order of this
phase transition. Lattice calculations by different groups do not agree 
on the order of this phase transition. 
Some lattice groups find the transition is second
order \cite{ke} and other groups find the transition is first order
\cite{giacomo}. Conventionally this transition is believed to be second 
order and in the universality class of $O(4)$ Heisenberg magnet 
\cite{rgopal}. But the effect of interaction 
between the chiral order parameter ($\Phi$) and the Polyakov loop ($L$)
may change the behavior of this transition. So in the present work
we investigate the effect of interaction between $\Phi$ and $L$ on
the 2-flavor chiral transition within an effective Lagrangian.

Previously, the effect of interaction between the chiral order parameter 
and the Polyakov loop has been studied in the 
renormalization group approach \cite{midorikawa}. The main difference 
between the present work and previous studies is that we consider the 
explicit breaking of the $Z_N$ symmetry. Explicit breaking of the 
$Z_N$ symmetry can be introduced in the effective Lagrangian by terms 
such as $\sim (L + L^\dag)$, $\sim (L + L^\dag)\Phi^\dag\Phi$.
Previous studies \cite{midorikawa} have considered interaction term,
such as $(LL^\dag)(\Phi^\dag\Phi)$, which respects both the chiral 
and the $Z_N$ symmetry. However, in the chiral limit, the
interaction terms need not respect the $Z_N$ symmetry. So terms such
as $\sim (L + L^\dag)\Phi^\dag\Phi$ should be considered. As we discuss 
later such a interaction term is always present if the explicit 
$Z_N$ symmetry breaking is large, for example in the
chiral limit. For simplicity we consider $N=2$ color QCD. We 
expect that different $N$ will not qualitatively change the physics 
we are discussing here. For $N_f=2$ the chiral order parameter $\Phi$ 
is a four component vector field whereas for $N=2$ the Polyakov
loop $L$ is a real scalar field. In this work we basically study 
the effect of the three terms, $~L$, $L \Phi^\dag\Phi$ and 
$L^2 \Phi^\dag\Phi$ in the effective Lagrangian. 
Our main result is that the strong explicit breaking of the
$Z_N$ symmetry
can make the chiral transition first order. We show that the chiral
transition can be first order even at the mean field level. We also
carry out numerical Monte Carlo simulations which confirm the first
order phase transition for large enough explicit $Z_N$ symmetry
breaking. We mention here that the $N_f=2$ chiral transition can be first
order from the interaction term $L^2\Phi^\dag\Phi$ without $Z_N$
symmetry breaking \cite{club}. However lattice QCD results indicate 
that effect of the $Z_N$ symmetric interaction term is small. A
possible first order chiral transition can result more likely from
the explicit $Z_N$ symmetry breaking as we will argue later.

Conventionally explicit
symmetry breaking weakens a phase transition. But our results suggest
that for a system with two order parameter fields
explicit symmetry breaking can make the transition stronger. It is
interesting to note that the chiral order parameter $\Phi$ does not couple
to gauge fields directly. The gauge fields seem to affect the
chiral phase transition indirectly, through the Polyakov loop. We
mention here that the effect of explicit symmetry breaking discussed here 
should not be restricted to the $Z_N$ symmetry. We expect that the explicit 
breaking of chiral symmetry may have some effect on the deconfinement 
transition in the large quark mass region. 
We mention here that interaction between the chiral order parameter and
the diquark fields is considered to study the chiral/color-superconducting
transition at low temperature and high density \cite{hatsuda}.

This paper is organized as follows. In section-II we describe the
effective Lagrangian and discuss the effect of the interaction between
$\Phi$ and $L$ on the chiral transtition. We describe our numerical
Monte Carlo calculations and results in section-III. The discussions and
conclusions are presented in section-IV.

\section{ The Effective Lagrangian and The Effect of The Broken
$Z_N$ Symmetry}

We consider the following Lagrangian \cite{sannino} in 3 dimension for 
the $\Phi$ and the $L$ fields,

\begin{eqnarray}
{\cal L} &=& {1\over 2}|\nabla\Phi|^2 + {1\over 2}(\nabla L)^2
+ V(\Phi,L),\nonumber\\
V(\Phi,L) &=& {m^2_\Phi\over 2}|\Phi|^2 + {\lambda_\Phi \over 4}|\Phi|^4 
+ {m^2_L\over 2}L^2 + {\lambda_L \over 4}L^4 
- g L^2|\Phi|^2 - cL|\Phi|^2 - eL
\end{eqnarray}

\noindent Some of the parameters of this reduced 3D Lagrangian depend
explicitly on temperature. This Lagrangian is invariant under 
$O(4)$ rotation of the $\Phi$ field. The last two terms of $V(\Phi,L)$
break the $Z_2$ symmetry ($L \to -L$) explicitly. 
The interactions between the chiral order parameter and the Polyakov
loop are taken care by 5th and 6th terms in $V(\Phi,L)$.

The signs of the couplings $g$ and $c$ decides the correlation between
the fluctuations of the $\Phi$ and the $L$ fields. For
example when $c>0$, the thermal average of the correlation between
the $\Phi$ and $L$ fluctuations, $\langle (\delta L)(\delta|\Phi|)\rangle$, is
positive. When $c<0$, $\langle (\delta L)(\delta|\Phi|)\rangle$ is negative. 
Such "anti-correlation" is seen between the fluctuations of the 
the chiral condensate and the Polyakov loop in the results of lattice 
QCD calculations\cite{dgl}. The correlation between the variations of the 
two order parameters with respect to temperature is of the same sign
as the correlation between the fluctuations. Note that
in the above Lagrangian when $c\ne 0$ the $\Phi$ field acts like an 
ordering field for the $L$ field. In the chiral symmetric phase the
chiral order parameter is small and the Polyakov loop expectation value
is large. The large expectation value of $L$, in the chiral symmetric phase,
can result only from the last term in $V(\Phi,L)$ (with $e>0$) 
because $|\Phi|$ is small.

\vskip0.2cm

As we have mentioned before previous studies \cite{midorikawa} have 
considered only the interaction term $gL^2|\Phi|^2$. When the coupling 
parameters $c=0=e$, the chiral transition and the deconfinement
transition do not always occur simultaneously. For large value 
of the coupling $g$ the transitions can occur simultaneously and
are of first order 
when coefficients of $|\Phi|^2$ and $L^2$ are negative in 
Eq.1 \cite{club,hatsuda}. A large positive $g$ would increase the 
critical temperature for the deconfinement transition as the coefficent
of $L^2$ term is negative for high temperatures. Lattice results on
the other hand show that inclusion of dynamical quarks decrease the
deconfinement transition temperature. So in QCD the coupling $g$ should 
be small. For small $g$ both the transitions are second order and
do not occur simultaneously. 

\vskip0.2cm

The coupling parameters $c$ and $e$ represent the strength of
the $Z_N$ explicit breaking. So they should increase with decreasing quark 
masses as the $Z_N$ breaking becomes severe. This can be seen explicitly
in the large quark mass region \cite{kg}. To see the effect
of the explicit $Z_N$ breaking let us consider $g=0=e$ and $c\ne 0$.
At the mean field level one can consider the temperature variation of
the parameters $m^2_\Phi$ and $m^2_L$. For simplicity
we fix $\lambda_{\phi,L} > 0$, $m^2_L>0$ and vary the $m^2_\Phi$ 
parameter. To find the $m^2_\Phi$ dependence of $L$ and $\Phi$ expectation
values one has to solve the following coupled equations,
\begin{eqnarray}
\lambda_\Phi |\Phi|^3 + m^2_\Phi|\Phi| - 2c|\Phi|L = 0\nonumber\\
\lambda_L L^3 + m^2_L L - c|\Phi|^2 = 0.\nonumber
\end{eqnarray}
We have checked numerically that for large enough $c$ these 
equations give two solutions which correspond to a degenerate minima
of the effective potential $V(\Phi,L)$ at some particular value of
$m^2_\Phi$. It may seem surprising that the potential $V(\Phi,L)$
has degenerate minima even though there is no cubic term for $\Phi$
and $L$ in it. However because of the coupling $c$ these two
fields are mixed. Though the mixing anle varies as
$m^2_\Phi$ is varied. With variation of $m^2_\Phi$ the minimum of
$V(\Phi,L)$, in the $|\Phi| - L$ plane, moves in directions other
than the $|\Phi|$ or $L$ axes. To understand how the minimum of $V(\Phi,L)$ 
behaves one should express $V(\Phi,L)$ in terms of variables which 
are the linear combinations of $|\Phi|$ and $L$. 
This would invariably lead to cubic
terms of the new fields in the effective potential. For some choice
of parameters the cubic term may then be important to cause degenerate
minima of $V(\Phi,L)$. Even if the explicit symmetry breaking is
not strong enough at mean field level fluctuations at higher order
can make the transition first order. At one loop 
the fluctuations of the $\Phi$ field will contribute to a non zero
3-point function of the $L$ field. This 3-point function can be
calculated perturbatively. In the high temperature approximation the 3-point 
function is given by,
\begin{equation}
\sim T{c^3 \over m^3},
\end{equation}
\noindent assuming zero momentum to all the external $L-$lines. $m$ here
is the mass of the $\Phi$ field fluctuations. Given a suitable choice
of parameters in the effective Lagrangian the three point function can
be significant. The consequence of this is a $\alpha(T)L^3$ term in the 
potential $V(\Phi,L)$ with temperature dependent $\alpha(T)$. This term
can cause discontinous change in $L$ as temperature varies. 
When $L$ field changes 
discontinuously the $\Phi$ field will also changes discontinuously because of 
the coupling term $cL\Phi^2$. If the coefficient of the $\Phi^2$ term 
changes from $+ve$ to $-ve$ due to discontinuous change in $L$ then $\Phi$ 
will change discontinuously from zero to non-zero. The $L^3$ term can also
come from other types of explicit symmetry breaking coupling terms but 
we think $cL\Phi^2$ is the simplest term in our model.\\ 

The situation with $c=0$ and $g\ne 0 \ne e$ is same as
the one discussed above. When the explicit symmetry breaking parameter $e$ 
is large, $Z_2$ symmetry of the $L$ field is lost and the $L$ field 
always has non-zero expectation value $L_0$. In order to study
the fluctuations one must expand 
the potential $V(\Phi,L)$ around $L_0$, $L=L_0 + \bar{L}$. This gives rise
to a term like $gL_0\bar{L}\Phi^2$ coming from the expansion of 
$gL^2\Phi^2$ around $L=L_0$. Now the term $gL_0\bar{L}\Phi^2$ is similar 
to the one discussed above. So there can be first order transition when 
$g \ne 0 \ne e$ like in the case when $c \ne 0$. We observed that 
at the mean field the transition becomes second order for large quartic 
couplings $\lambda_\Phi$. Now in the following
section we discuss the numerical Monte Carlo simulations of the effective
Lagrangian (Eq.1) and the results. 
These calculations include higher order as well as non perturbative 
fluctuations.

\section{Numerical calculations and results}

The numerical Monte Carlo calculations of the the model Eq.1 is done by 
discretizing the action $S=\int {\cal L}d^3x$ on a 3 dimensional $N_s^3$
lattice. We employ the following rescaling of the fields variables and
the couplings,
\begin{eqnarray}
&\Phi& \to {\Phi\sqrt{\kappa_\Phi} \over a}, \lambda_\Phi \to
{\lambda_\Phi \over \kappa^2_\Phi}, m^2_\Phi \to {2-4\lambda_\Phi -
6\kappa_\Phi \over \kappa_\Phi a^2},\nonumber\\
&L& \to {L\sqrt{\kappa_L} \over a}, \lambda_L \to
{\lambda_L \over \kappa^2_L}, m^2_L \to {2-4\lambda_L -
6\kappa_L \over \kappa_L a^2},\nonumber\\
&g& \to {g \over \kappa_\Phi\kappa_L}, c \to {c \over a \sqrt{\kappa_L}
\kappa_\Phi}, e \to {e\over a^3 \sqrt{\kappa_L}}.
\end{eqnarray}
\noindent $a$ is the lattice spacing and $\kappa_\Phi$ and $\kappa_L$
are the hopping parameters for the $\Phi$ and the $L$ fields respectively. 
After the rescaling of the fields and the coupling parameters the 
discretized lattice action becomes,
\begin{eqnarray}
S=\sum_x &-&\kappa_\Phi\sum_\mu \Phi_x\Phi_{x+\mu} + |\Phi_x|^2 + 
\lambda_{\Phi}(|\Phi_x|^2 - 1)^2 \nonumber\\
&-&\kappa_L\sum_\mu L_xL_{x+\mu} + L^2_x + 
\lambda_{L}(L^2_x - 1)^2 \nonumber\\
&-&gL^2_x|\Phi_x|^2-cL_x|\Phi_x|^2-cL_x
\end{eqnarray}
\noindent Here $\Phi_x(L_x)$ represents the value of the $\Phi(L)$ field
at the lattice site $x$. $x+\mu$ represents the six nearest neighbor
lattice sites to $x$. We adopted the pseudo heat-bath method used for 
the Higgs updating in SU(2)+Higgs studies \cite{bunk}. To update $\Phi_x$ 
at a lattice site $x$ we write the probability distribution $P(\Phi_x)$ 
of $\Phi_x$ as,
\begin{eqnarray}
P(\Phi_x) &\sim& Exp\left[-S_1(\Phi_x)-S_2(\Phi_x)\right],~~~ \rm{with}\nonumber \\
S_1(\Phi_x) &=& \alpha\left(\Phi_x - {A\over 2\alpha}\right)^2,~S_2(\Phi_x) =
\lambda_{\Phi}\left(|\Phi_x|^2 - B\right)^2\nonumber\\
A &=& \kappa_\Phi \sum_\mu \Phi_{x+\mu},~~
B = 1 - {1 \over 2\lambda_\Phi} + {\alpha\over 2\lambda_\Phi} 
+ {cL_x + gL_x^2\over 2\lambda_\Phi}
\end{eqnarray}
The coefficient $\alpha$ is a parameter chosen so that we get a
reasonable acceptance rate for the new $\Phi_x$. Once $\alpha$ is chosen,
a Gaussian random number is generated according to the distribution
Exp$\left[-S_1(\Phi_x)\right]$. Then this random number is accepted as the
new value of $\Phi_x$ with the probability Exp$\left[-S_2(\Phi_x)\right]$.
Using this procedure $\Phi_x$ is updated at all the lattice sites. 
Then we do the updating of $L_x$ along the same 
steps. The process of updating $\Phi_x$ and 
$L_x$ on the entire lattice is repeated about 20 times between
successive measurements. We measure the magnetizations 
\begin{equation}
M_\Phi = {1\over V}\sum_x\Phi_x,~~~~ M_L = {1\over V}\sum_xL_x, 
\end{equation}
\noindent where $V = N_s^3$ is the number of lattice sites. The 
expectation values of the $\Phi$ and the $L$ fields are given by the thermal 
averages (average over the measurements), 
$\left<\Phi\right> = \left< |M_{\Phi}|\right>$ and 
$\left<L\right> = \left< M_{L}\right>$. We take the absolute value 
of $M_\Phi$ for $\left<\Phi\right>$ because a 
normal average of $M_\Phi$ is usually not a well behaved observable. 
\vskip0.2cm
The numerical simulations were carried out on a $N_s=16$ lattice. In this
work we do not intent to explore the phase diagram of the model (Eq.1) but
to show that for suitable choice of parameters the phase transition can 
change from second order to first order.
Here we present results for two sets of parameters. For one set we fix the 
couplings $g=e=0$ and for the second set we fix the coupling $c=0$. 
For the first set of parameters we choose, $\lambda_{\Phi}=0.004$, 
$\lambda_{L}=0.0020$, $\kappa_L = 0.01$ and $c=0.1$. We observe the
hysteresis of $\left<\Phi\right>$ and $\left<L\right>$ by varying the 
parameter $\kappa_\Phi$. In FIG.1 we show the hysteresis loop of 
$\left<\Phi\right>$ and in FIG.2 we show the hysteresis loop of 
$\left<L\right>$.

\begin{figure}[t!]
\begin{center}
\begin{minipage}{.45\textwidth}
\includegraphics[width=7.cm]{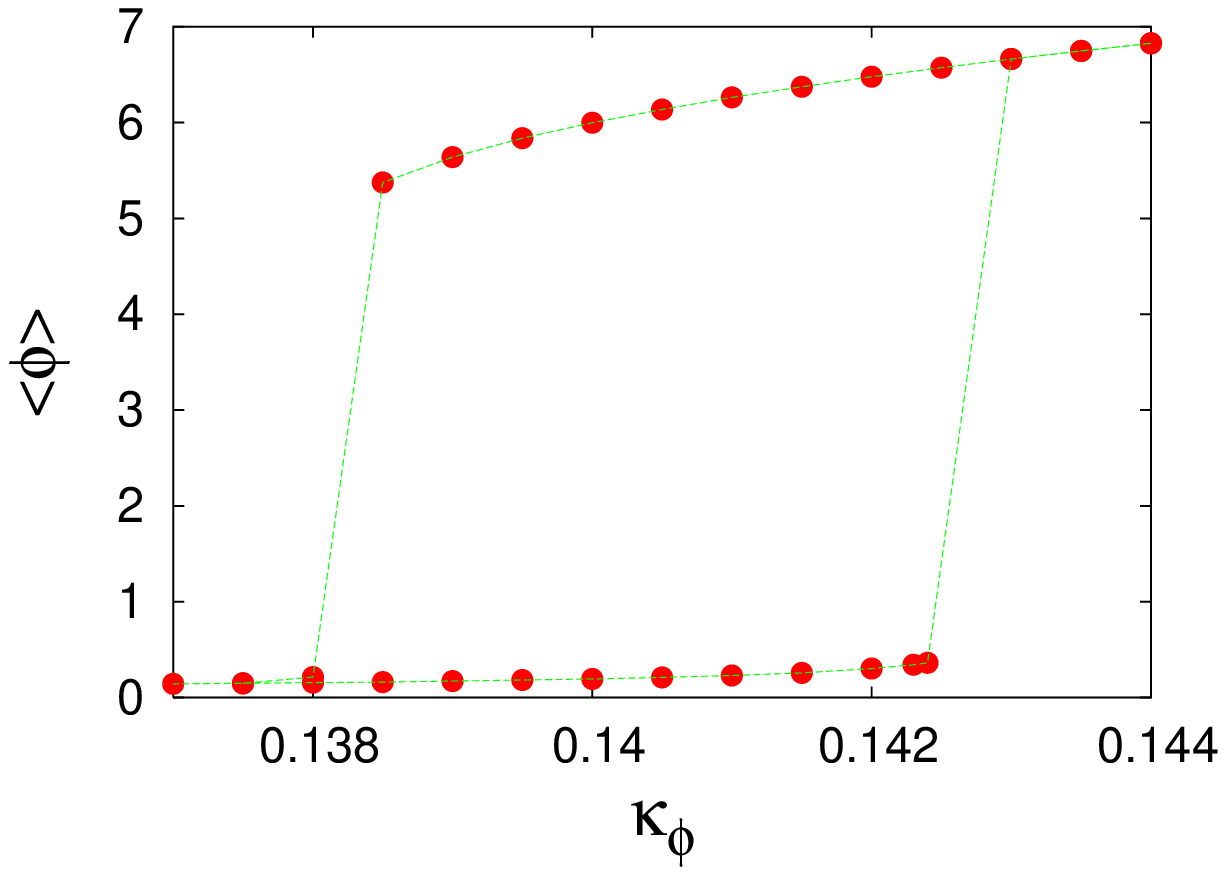} \label{mf}
\caption{The hysteresis of $\left<\Phi\right>$ vs $\kappa_\Phi$}
\vskip -0.3cm
\end{minipage}
\hspace{3.5em}
\begin{minipage}{.45\textwidth}
\includegraphics[width=7.cm]{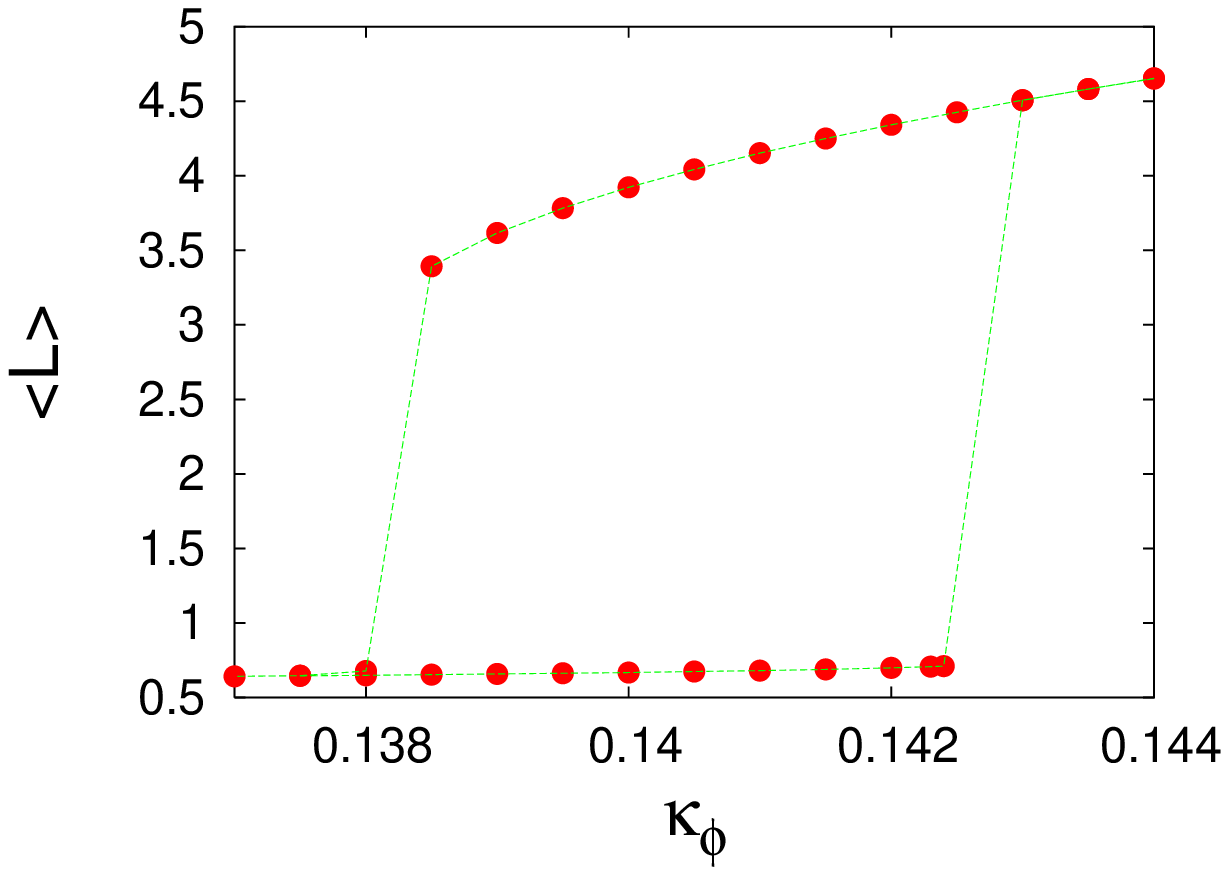}  \label{pt}
\vskip -0.3cm
\caption{The hysteresis of $\left<L\right>$ vs $\kappa_\Phi$}
\vskip -0.3cm
\end{minipage}
\end{center}
\end{figure}

Since we take $c$ to be positive we see $\left<\Phi\right>$ and 
$\left<L\right>$ increase or decrease simultaneously. The choice of the 
values for the parameters is such that the variation of $\left<\Phi\right>$ and
$\left<L\right>$ are similar in magnitude. For $c < 0$ increase in 
$\left<\Phi\right>$ should lead to decrease in $\left<L\right>$. So in 
this case the hysteresis loop for $\left<\Phi\right>$ will look somewhat
similar to Fig.1 while the hysteresis loop for $\left<L\right>$ will
be more or less inverted about the y-axis. 
\vskip0.2cm
For the second set of 
parameters we choose $\lambda_{\Phi}=0.0055$, $\lambda_{L}=0.0010$, 
$\kappa_L=0.14$, $g=-0.02$ and $e=0.9$. The hysteresis loop of 
$\left<\Phi\right>$ and $\left<L\right>$ are observed by again varying the 
parameter $\kappa_\Phi$. The choice of $\kappa_L$ and $e$ is such that 
the expectation value of $L$ is non-zero and positive. The sign of $g$ 
assures that increase in $\left<\Phi\right>$ leads to decrease in 
$\left<L\right>$ and vice-versa as the parameter $\kappa_\Phi$ is varied. 
The hysteresis curves of the two
order parameters are shown in FIG.3 and FIG.4.
The values of $\lambda_{\Phi,L}$ in our calculations are chosen 
so that we can see first order phase transition clearly and the variation of 
$\left<\Phi\right>$ and $\left<L\right>$ are of order O(1) across the 
transition point. Note that with suitable choice of $\kappa_L$ and $\lambda_L$
one change the average of the Polyakov loop across the transition point.
The choice $g$ was such that the chiral transition turned out to be 
second order when the coupling $e$ was set to zero.
We also did simulations on a 4D lattice. The results in
this case are very similar to the 3D simulations.

\begin{figure}[h!]
\begin{center}
\hskip-1.3cm
\begin{minipage}{.45\textwidth}
\includegraphics[width=8.cm]{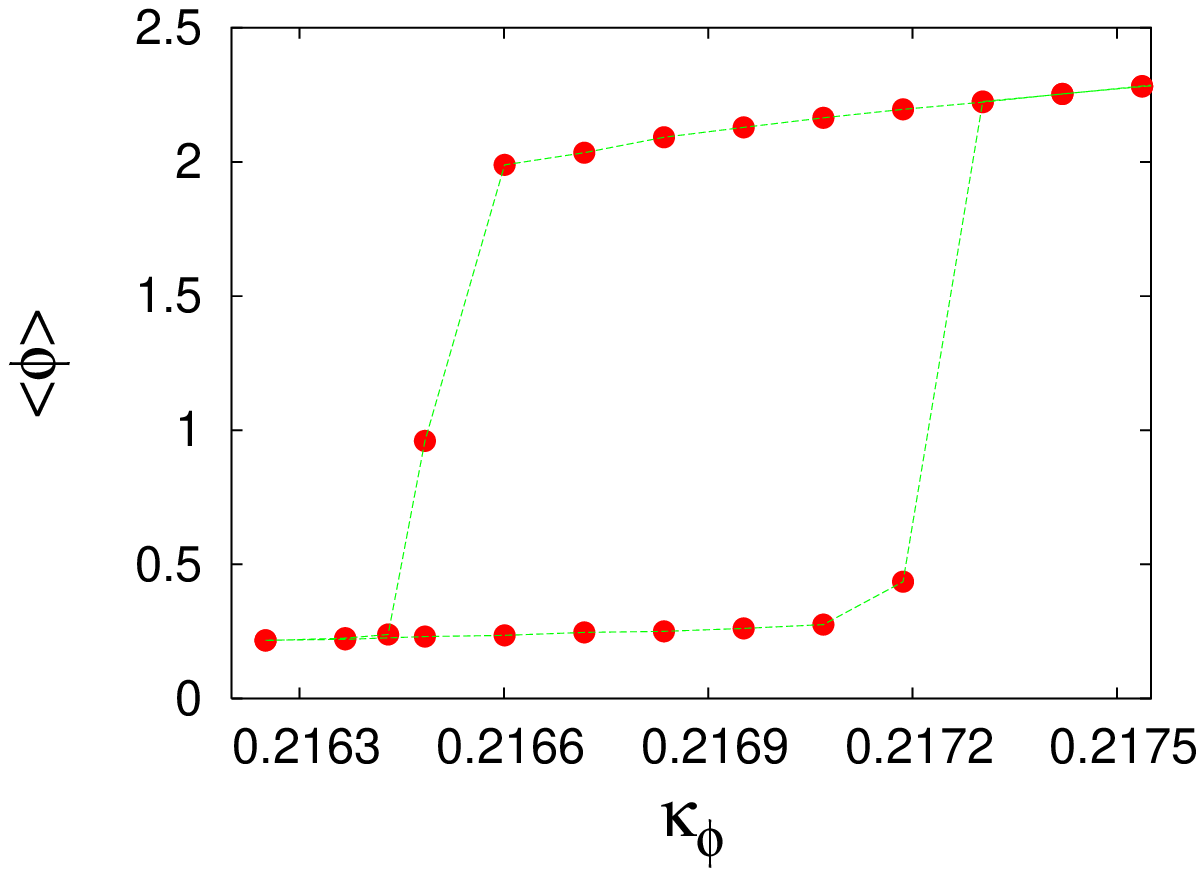} \label{mf}
\vskip -0.3cm
\caption{The hysteresis of $\left<\Phi\right>$ vs $\kappa_\Phi$}
\vskip -0.3cm
\end{minipage}
\hspace{3.5em}
\begin{minipage}{.45\textwidth}
\includegraphics[width=8.cm]{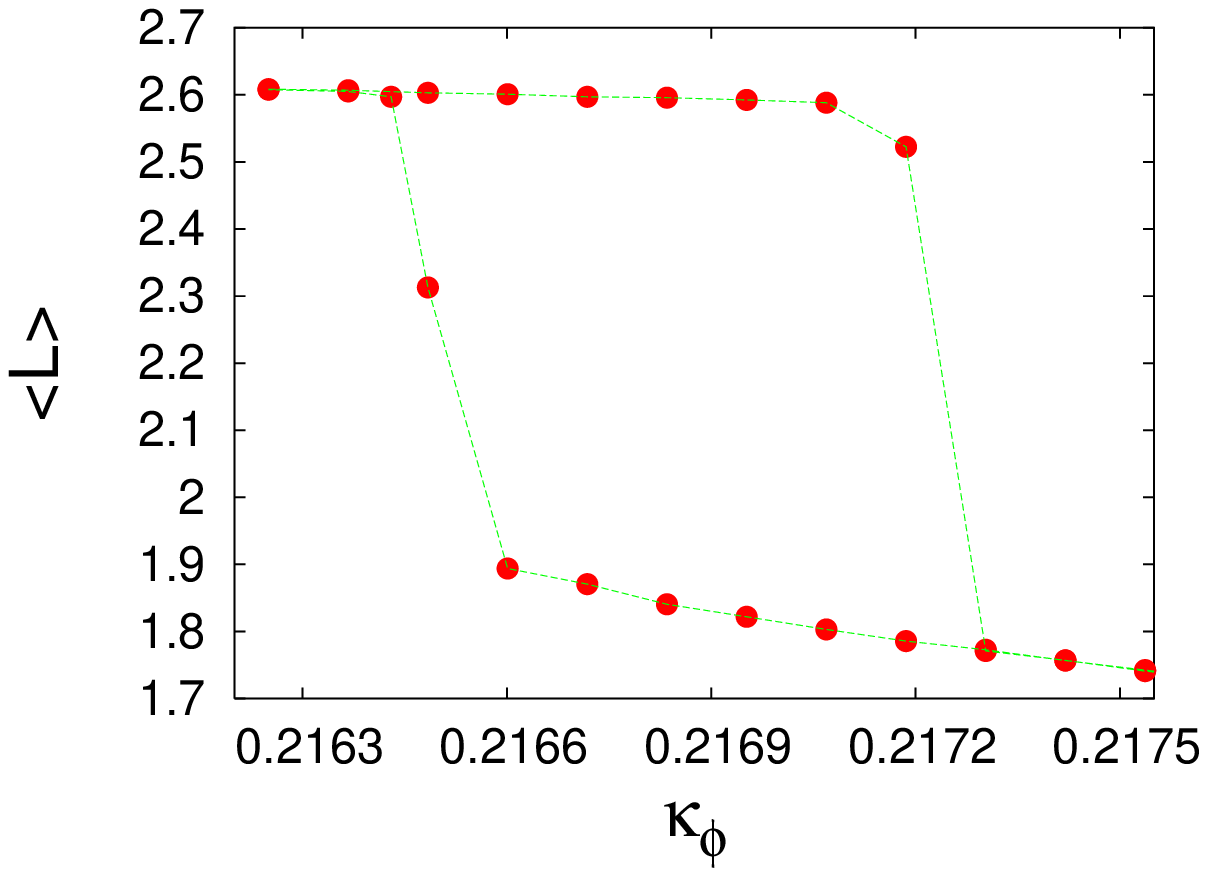}  \label{pt}
\vskip -0.3cm
\caption{The hysteresis of $\left<L\right>$ vs $\kappa_\Phi$}
\vskip -0.3cm
\end{minipage}
\end{center}
\end{figure}

The results in FIG.1-4 show strong first order phase transition for the $\Phi$ 
and $L$ fields. By fixing the coupling parameters $g$, $c$ and $e$ we
observed that the strength of the transition depends on the values of
the quartic couplings $\lambda_\Phi$ and $\lambda_L$. The transition
becomes weaker with increase in any of the quartic couplings 
$\lambda_{\Phi,L}$.
However even for larger quartic couplings a suitable choice of the
parameters $g$, $c$ and $e$ made the transition strong first order.

\vskip0.2cm
We also did calculations with small explicit symmetry breaking for
the $\Phi$ field by considering a linear $\Phi$ term in the Lagrangian.
In this case we found that the hysteresis loops of both
$\left<\Phi\right>$ and $\left<L\right>$ shrinking in size with increase
in the coefficient of linear $\Phi$ term in the Lagrangian. These results 
suggest that the transition becomes weaker when the chiral symmetry is 
explicitly broken.

\section{Discussion and Conclusions}

Using a simple effective Lagrangian, which captures the chiral symmetry
and $Z_N$ symmetry of QCD, we have investigated the effect of the explicit
symmetry breaking on the chiral phase transition. Since we consider the
chiral limit the $Z_2$ symmetry of the Polyakov loop is explicitly
broken. As we incorporate the explicit $Z_2$ breaking interaction terms in the 
effective Lagrangian we find the chiral transition becomes first
order. We observed that the first order transition becomes weaker when a 
small explicit symmetry breaking is considered for the chiral order
parameter $\Phi$. The results of our calculations show that two flavor 
chiral transition is of first order for large enough $Z_N$ explicit breaking. 
As we have mentioned before at present some lattice studies suggest that the 
transition is first order \cite{giacomo} and some other studies show
the transition is second order \cite{ke}. These conflicting 
results may be because the quark masses studied are not small enough or
the lattices used are not big enough.

In the previous study, for $N_f=2$ and $N=3$ \cite{midorikawa}, the
second order chiral transition becomes first order when coupled to the 
Polyakov loop. This is because the deconfinement transition is first
order for $N=3$ with a cubic term $\sim(L^3 + L^{*3})$ in the effective
potential with exact $Z_3$ symmetry. However when the quark masses
are finite and decrease the deconfinement transition becomes weaker. This
can be understood due to explicit breaking of $Z_3$ symmetry. Already in 
the large quark mass region the deconfinement transition becomes
crossover which implies the the explicit symmetry breaking dominates over
the effects of the above $Z_3$ symmetric cubic term.
For smaller quark masses the explicit $Z_3$ symmetry breaking likely grows
and expected to be maximal in the chiral limit. So for smaller quark 
masses, i.e in the chiral limit one should rather consider the effect of 
the explicit breaking of the $Z_3$ symmetry. The effects of explicit
symmetry breaking discussed in this work should not be restricted
to the explicit breaking of the $Z_N$ symmetry. For 2-flavor and 2-color
QCD the deconfinement transition in the large quark mass region may 
have the effects coming from the explicit breaking of the chiral symmetry.
It may be possible that this effect change the phase transition behavior
of the deconfinement transition in the heavy quark mass region.

{\bf Acknowledgments}

\medskip

We have benefited greatly from discussions with T. Hatsuda and 
A. M. Srivastava. We would like to thank G. Baym, S. Datta, M. Laine, 
T. Matsuura and M. Ohtani for comments. This work is supported by the JSPS
Postdoctoral Fellowship for Foreign Researchers.

\end{document}